\documentclass[         %
aps,                    
prd,                    
showpacs,               
nofootinbib,            
showkeys,               %
preprintnumbers,        %
floatfix]               
{revtex4}               
\usepackage{graphicx,longtable}

\begin{document}

\title{Constraints on Unparticles from Low Energy Neutrino-Electron 
Scattering}

\author{A. B. Balantekin}\email{baha@physics.wisc.edu}
\affiliation{Department of Physics, University of Wisconsin, Madison, 
WI 53706, USA}
\author{K. O. Ozansoy}\email{oozansoy@physics.wisc.edu}
\affiliation{Department of Physics, University of Wisconsin, Madison, 
WI 53706, USA}
\affiliation{Department of Physics, Ankara University, 06100 Tandogan, 
Ankara, Turkey}

\date{\today}

\begin{abstract}
We place limits on scalar and vector unparticle couplings to electrons 
and neutrinos 
using data from reactor neutrino experiments originally designed to 
search for neutrino magnetic moment. 
Upper bounds on Standard Model lepton-scalar unparticle and lepton -vector 
unparticle couplings $\lambda_0$,
and $\lambda_1$
are given for various values of the unparticle mass dimension $d$, and  
the unparticle energy scale $\Lambda_{\cal U}$.
Especially for smaller values of the mass dimension $d$ ($1 <d <1.3$) 
these bounds are very significant and comparable to those obtained from 
production rates at high energy colliders. These bounds are also 
similar to those 
obtained by the analysis of the absence of the decay of the low-energy 
solar and reactor neutrinos. 
\end{abstract}
\medskip


\maketitle

\section{Introduction}

Georgi suggested that a scale invariant sector which decouples at a 
sufficiently large energy scale may exist \cite{Georgi:2007ek}. Presence 
of this sector could yield very interesting physics beyond the 
Standard Model and should be accessible at colliders. 
If such a scale invariance occurs in nature, it cannot be described in 
terms of 
particles. In this scheme the hidden sector with a non-trivial 
infrared fixed point, 
described by the Banks-Zaks (BZ) \cite{Banks:1981nn} operators 
${\cal O}_{BZ}$ and the Standard Model sector, described by the 
operators ${\cal O}_{SM}$ 
interact through the exchange of
particles with a mass scale of ${\cal M}_{\cal U}^k$: 
\begin{equation} 
\label{eqn:eq1} 
 \frac{1}{{\cal M}_{\cal U}^k}{O}_{BZ}{O}_{SM}, 
\end{equation}
where BZ and Standard Model operators have the mass dimension $d_{BZ}$ 
and $d_{SM}$, respectively.  
Below the infrared fixed point $\Lambda_{\cal U}$, the BZ operators 
${\cal O}_{BZ}$ mutate into unparticle operators and 
Eq. (\ref{eqn:eq1}) reduces to  
\begin{equation}
\label{eqn:eq2}
 \frac{C_{\cal U} 
\Lambda_{\cal U}^{d_{BZ}-d}}{{\cal M}_{\cal U}^k}{O}_{BZ}{O}_{SM}, 
\end{equation}
where $d$ is the non-integral scaling mass dimension of the unparticle 
operator 
$O_{\cal U}$, and the constant $C_{\cal U}$ is the coefficient of the 
relevant operators. (Note that in Ref. \cite{Georgi:2007ek}
the scale dimension is defined by $d_{\cal U}$).

Following Georgi's original paper there has been significant 
activity investigating phenomenological 
consequences of the unparticle sector. 
Implications of the interference of the Standard Model amplitudes 
and amplitudes with 
virtual unparticles were considered in \cite{Georgi:2007si}. 
Collider phenomenology of unparticle physics has been explored in a great 
detail, and Feynman rules for spin 0, spin 1, or spin 2 unparticles coupled 
to a variety of Standard Model gauge invariant operators 
have been explicitly given in Ref. 
\cite{Cheung:2007ue}. Subsequently many authors studied possible signatures 
of unparticle sector in collider experiments 
\cite{Rizzo:2007xr,Bander:2007nd,Chen:2007qr,Zhou:2007zq,Luo:2007bq,Aliev:2007gr}. 
Although the small magnitude of the coupling between 
unparticles and Standard Model 
could conceal the unparticle phenomena at low energies,  
unparticle physics can nevertheless be constrained using data from 
non-accelerator physics. Limits have indeed 
been placed on unparticle - Standard 
Model particle couplings 
from cosmology, especially big-bang nucleosynthesis \cite{Davoudiasl:2007jr}, 
from avoidance of supernova overcooling by unparticle emission 
\cite{Davoudiasl:2007jr,Hannestad:2007ys,Freitas:2007ip}, 
and from the limits on the 
decay of solar and reactor neutrinos in oscillation experiments 
\cite{Anchordoqui:2007dp}. 

In this paper, we consider a complementary  non-accelerator limit on 
unparticle-Standard Model 
couplings, namely very low-energy elastic neutrino-electron scattering 
experiments. Typically such experiments are performed at sources that 
emit a large number of neutrinos such as reactors and their search for 
physics beyond Standard Model is characterized by the neutrino magnetic 
moment. 
(For a recent review see e.g. Ref \cite{Balantekin:2006sw}). Nevertheless, 
the data from such experiments can also be used 
to search for other physics beyond 
the Standard Model. 
In the next section we place limits on scalar and vector unparticle 
couplings to electrons and neutrinos 
using data from reactor neutrino experiments originally designed to 
search for neutrino magnetic moment.
We find that our limits are comparable to those 
obtained from the decay bounds of solar and reactor neutrinos.

\section{Antineutrino-Electron Scattering at Reactors}

The differential Standard Model cross section for electron antineutrinos on 
electrons is given by \cite{Vogel:1989iv}
\begin{equation}
\label{3}
 \frac{d\sigma}{dT}=\frac{G_F ^2 m_e}{2\pi}
\left[ (g_V+g_A)^2+(g_V-g_A)^2 \left( 1-\frac{T}{E_\nu} \right)^2
+[g_A^2-g_V^2]\frac{m_e T}{E_\nu} \right], 
\end{equation}
where $T$ is the electron recoil kinetic energy, 
$g_V=2\sin^2 \theta_W +\frac{1}{2}$, $g_A=\frac{1}{2}$ for $\nu_e$,
$g_A=-\frac{1}{2}$ for $\bar \nu_e$. Low-energy antineutrino-electron elastic 
scattering experiments typically search for excess counts beyond the 
Standard Model contribution given in Eq. (\ref{3}). 

Interactions between Standard Model leptons and the scalar unparticles are 
given by \cite{Georgi:2007si,Cheung:2007ue,Zhou:2007zq}
\begin{equation}
\label{4}
\lambda_{0e}\frac{1}{\Lambda^{d-1}}\bar e e O_{\cal U} +
\lambda_{0\nu}^{\alpha\beta}\frac{1}{\Lambda^{d-1}} \bar 
\nu^{\alpha} \nu^{\beta}O_{\cal U}+ {\rm h.c.} . 
\end{equation}
Therefore, the contribution to the scattering amplitude 
for elastic $\bar \nu_e$-$e$ scattering from the exchange of the scalar 
unparticle takes the form 
\begin{equation}
\label{5}
{\cal M}_{{\cal U}_S}=\frac{f(d)}{\Lambda_{\cal U}^{2d-2}} 
[\bar \nu_\beta(k') \nu_\alpha(k)][\bar e(p') e(p)][-q^2-i\epsilon]^{d-2}, 
\end{equation}
where 
\begin{equation}
\label{6}
f(d)=\frac{{\lambda_{0\nu}}^{\alpha\beta}\lambda_{0e} A_d}{2 \sin (d\pi)},
\end{equation}
and 
\begin{equation}
\label{7}
A_d=\frac{16\pi^{5/2}}{{(2\pi)}^{2d}} 
\frac{\Gamma(d+1/2)}{\Gamma(d-1)\Gamma(2d)}. 
\end{equation} 
In writing Eqs. (\ref{5}) and (\ref{6}) we included the possibility that 
scalar unparticle exchange may change the neutrino flavor. 
In our analysis for scalar unparticles, we use the short-hand notation $
\lambda_0 \equiv (\lambda_{0\nu}\lambda_{0e})^{1/2}$. 
The interference term 
between the scalar unparticle amplitude and the Standard Model amplitude
is proportional to the neutrino mass (or more precisely to 
$m_{\nu}/\Lambda$) and we neglect its contribution to the
cross section. The contribution to the differential scattering cross section 
from the exchange of the scalar unparticle then takes the form 
\begin{equation}
\label{8}
\frac{d\sigma}{d T}=
\frac{f(d)^2(2^{(2d-7)})}{\pi {E_\nu}^2 \Lambda_{\cal U}^{4d-4}} 
(m_e T)^{(2d-3)}(T+2m_e) .
\end{equation}
For small electron recoil energies ($T < 2m_e$), this cross-section 
depends on T as $1/T^{(3-2d)}$. Hence as $T$ gets smaller the differential 
cross section gets larger for $d<3/2$. This behavior is reminiscent 
of the T-dependence of the neutrino magnetic moment contribution, which 
goes like $\sigma \sim \mu^2/T$. If d were allowed to take the value $d=1$, 
the cross section in Eq. (\ref{8}) would behave just like the magnetic moment 
cross section. For this reason experiments designed to search for neutrino 
magnetic moment are particularly appropriate to place limits on 
unparticle couplings, especially for values of $d$ close to 1. 

Interactions between Standard Model leptons and the vector unparticles are 
given by \cite{Georgi:2007si,Cheung:2007ue,Zhou:2007zq}
\begin{eqnarray}
\label{9}
\lambda_{1e}\frac{1}{\Lambda^{d-1}}\bar e \gamma_\mu e O_{\cal U}^\mu +
\lambda_{1\nu}^{\alpha\beta}\frac{1}{\Lambda^{d-1}} 
\bar \nu^{\alpha} \gamma_\mu \nu^{\beta}O_{\cal U}^\mu+h.c. 
\end{eqnarray}
Hence the contribution to the scattering amplitude 
for elastic $\bar \nu_e$-$e$ scattering from the exchange of the vector 
unparticle takes the form 
\begin{eqnarray}
\label{10}
{\cal M}_{{\cal U}_V}=\frac{f(d)}{\Lambda_{\cal U}^{2d-2}} 
[\bar \nu_\beta(k') \gamma_\mu \nu_\alpha(k)][\bar e(p') \gamma_\nu e(p)]
[-P^2-i\epsilon]^{d-2}\pi^{\mu\nu}(P), 
\end{eqnarray}
where 
\begin{eqnarray}
\label{11}
\pi_{\mu\nu}(P)=&&-g^{\mu\nu}+\frac{P^\mu P^\nu}{P^2}.
\end{eqnarray}
Similar to the scalar unparticle case we use 
the short-hand notation 
$\lambda_1 \equiv (\lambda_{1\nu}\lambda_{1e})^{1/2}$. 
The contribution to the differential scattering cross section 
from the exchange of the vector unparticle then takes the form
\begin{eqnarray}
\label{12}
\frac{d\sigma}{d T}=
\frac{(f(d))^2(2)^{(2d-8)}}{\pi \Lambda_{\cal U}^{4d-4}}
(m_e)^{2d-3} (T)^{(2d-4)}\big[ 1+\big(1-\frac{T}{E_\nu}\big)^2 - 
\frac{m_eT}{E_\nu^2}\big]
\end{eqnarray}

\begin{figure}
\includegraphics{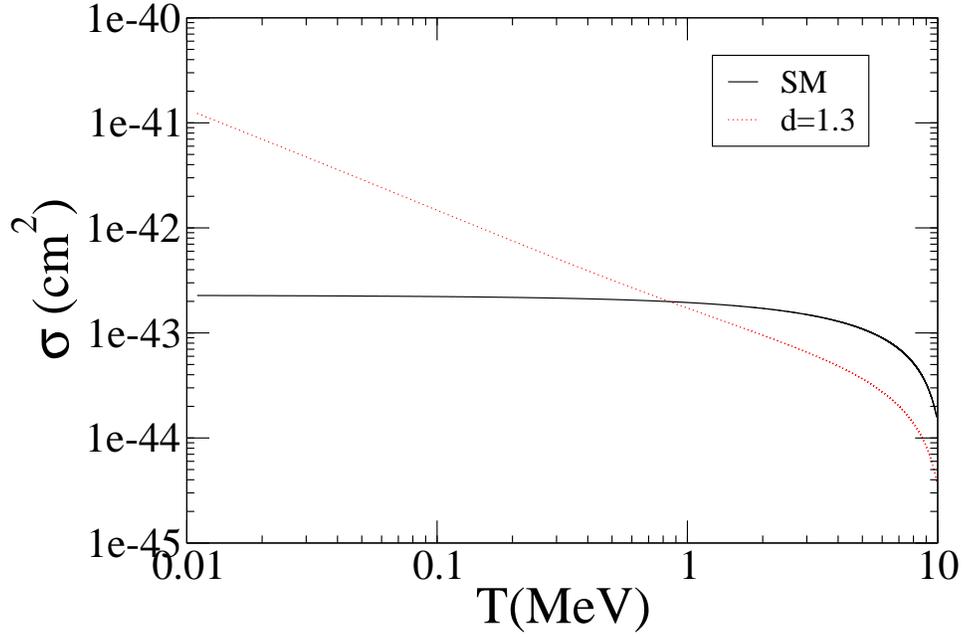}
\caption{\label{fig:ex}
Folded differential cross sections from the Standard Model and scalar 
unparticles. 
Solid curve is for the Standard Model and the dashed one for the 
scalar unparticle.
We assumed $d=1.3$, $\Lambda_{\cal U} = 1$ TeV, and 
$\lambda_{0 {\rm max}}=0.001$.} 
\end{figure}

Nuclear reactors emit copious ($\sim 10^{20}$ per second) electron 
antineutrinos with energies up to about 10 MeV. 
The exact energy spectrum depends on the specific fuel composition of 
the reactor. As an example, reactor neutrino spectrum 
coming from the fissioning of $^{235}U$ is approximately given by 
\cite{Vogel:1989iv}
\begin{equation}
\label{13}
  \frac{dN_\nu}{dE_\nu} \sim \exp{(0.870-0.160E_\nu-0.0910E_\nu^2)},
 \end{equation}
where the antineutrino energy, $E_{\nu}$, is given in MeV. 
The pertinent quantity for these elastic scattering experiments is 
the folded cross-section:
\begin{equation}
\label{14}
\left\langle \frac{d\sigma}{dT} \right\rangle = 
\int_{E^{\text{min}}_{\nu}(T)}^\infty
\frac{dN_\nu}{dE_\nu}\frac{d\sigma(E_\nu)}{dT}dE_\nu, 
\end{equation}
where $E_\nu^{\text{min}}(T)=0.5(T+\sqrt{T^2+2Tm_e})$.
In Figure \ref{fig:ex} we contrast schematic behaviors of the 
Standard Model cross section and the cross section from the scalar unparticle 
exchange 
folded with the reactor spectrum. 
In the calculations leading to this figure we assumed that $d=1.3$, 
$\Lambda_{\cal U} = 1$ TeV, and  
$\lambda_{0 {\rm max}}=10^{-3}$. 
The figure indicates that at low electron recoil energies the unparticle 
contribution could measurably change the total count rate. As in 
the neutrino magnetic moment searches, the crucial quantity is the minimum 
electron recoil energy accessible to the experiment. 
Current experimental status of neutrino magnetic moment searches 
is summarized by the Particle Data Group 
\cite{Yao:2006px}. Two recent experiments are the MUNU experiment 
with an energy threshold of 700 KeV \cite{Daraktchieva:2005kn} and 
the TEXONO experiment with a threshold of 5 KeV \cite{Wong:2006nx}. 
\begin{table} [b]
\caption{ \label{tab:d-l1} 
For $\Lambda_{\cal U}=1$ TeV, upper limits of coupling 
constant $\lambda_0$ for 
various values of mass dimension d.}
\begin{center}
\begin{tabular}{|c|c|}
\hline
 $d$ & $\lambda_{0max}$ \\
\hline
1.01 & $3.5\times 10^{-6}$\\
\hline
1.05 & $7.3\times 10^{-6}$\\
\hline
1.1 & $1.9\times 10^{-5}$\\
\hline
1.2 & $1.2\times 10^{-4}$\\
\hline
1.3 & $7.2\times 10^{-4}$\\
\hline
1.4 & $4.5\times 10^{-3}$\\
\hline
1.5 & $2.7\times 10^{-2}$\\
\hline
1.7 & $9.5\times 10^{-1}$\\
\hline
1.9 & $24.5 $\\
\hline
\end{tabular}
\end{center}
\end{table}

\begin{table} [b]
\caption{ \label{tab:d-l2} 
For $\Lambda_{\cal U}=1$ TeV, upper limits of coupling constant 
$\lambda_1$ for 
various values of mass dimension d.}
\begin{center}
\begin{tabular}{|c|c|}
\hline
 $d$ & $\lambda_{1max}$ \\
\hline
1.01 & $1.1\times 10^{-6}$\\
\hline
1.05 & $2.4\times 10^{-6}$\\
\hline
1.1 & $6.2\times 10^{-6}$\\
\hline
1.2 & $3.8\times 10^{-5}$\\
\hline
1.3 & $2.0\times 10^{-4}$\\
\hline
1.4 & $1.4\times 10^{-3}$\\
\hline
1.5 & $9.1\times 10^{-3}$\\
\hline
1.7 & $3.1\times 10^{-1}$\\
\hline
1.9 & $8.1$\\
\hline
\end{tabular}
\end{center}
\end{table}

\begin{figure}
\includegraphics{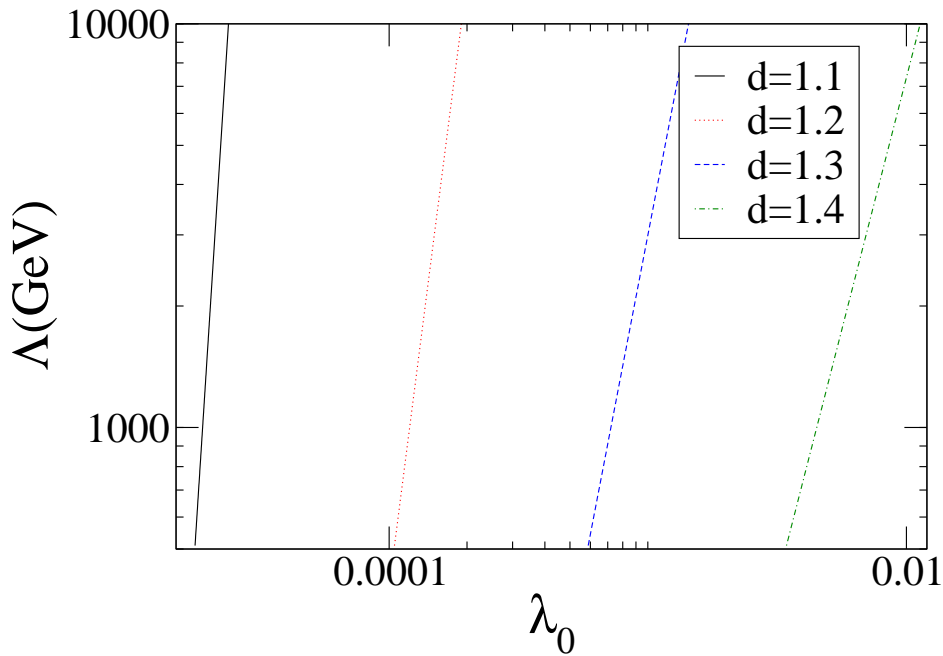}
\caption{\label{fig:l-l}
Upper limits on the maximum value of the scalar unparticle 
coupling parameter $\lambda_0$ for 
different energy scales $\Lambda_{\cal U}$ obtained using the TEXONO 
data.}
\end{figure}

\begin{figure}
\includegraphics{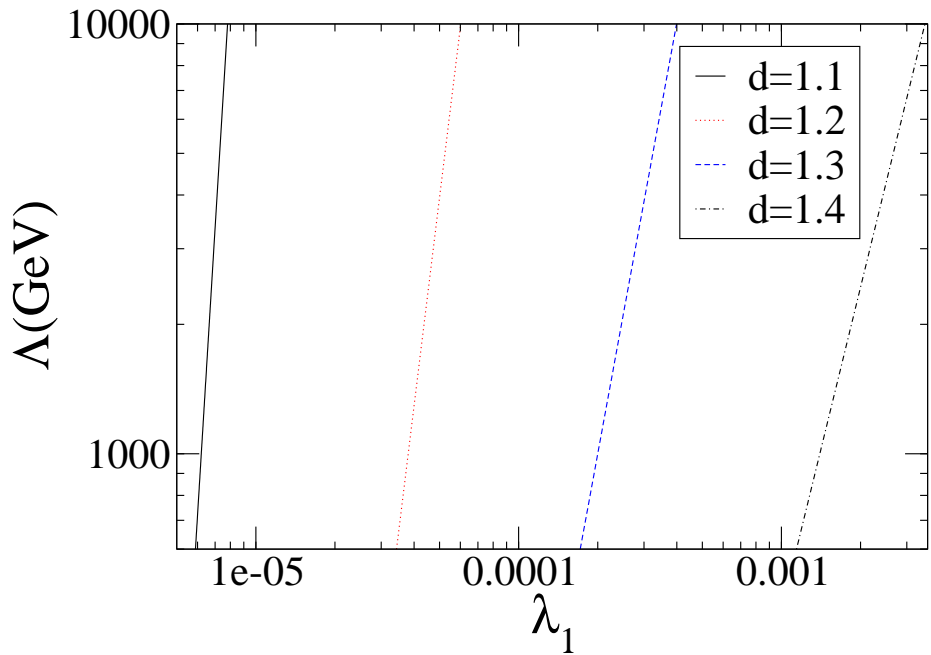}
\caption{\label{fig:l-l2}
Upper limits on the maximum value of the vector unparticle 
coupling parameter $\lambda_1$ for 
different energy scales $\Lambda_{\cal U}$ obtained using the TEXONO 
data.}
\end{figure}

In our calculations, we fix the unparticle mass dimension 
$d$ to specific values. Upper limits on the coupling constants 
$\lambda_0$, and $\lambda_1$ 
with $\Lambda_{\cal U}=1$ TeV for various values of the mass dimension $d$, 
obtained from the analysis of the TEXONO data are shown in 
Table \ref{tab:d-l1}, and  Table \ref{tab:d-l2}.
Future experiments are expected to try lowering the energy threshold. 
Figure \ref{fig:ex} clearly indicates that, if lower energy thresholds can be 
achieved, limits on the coupling constants can be significantly tightened. 
Since the cross section is proportional 
to the combination  
$\lambda^4/\Lambda_{\cal U}^{4d-4}$, in principle it is this quantity that 
can be extracted from the low-energy elastic electron-antineutrino 
scattering data.
The resulting scaling behavior for the scalar and vector unparticle couplings
are illustrated in Figures \ref{fig:l-l} and \ref{fig:l-l2}, respectively.  
In these figures, the parameter space to the right of a particular line is 
ruled out by the TEXONO data for the indicated value of $d$.

\section{Conclusions}

We obtained bounds on the coupling of the unparticle sector to electron 
and electron neutrinos using elastic scattering data from reactors. 
Especially for smaller values of the mass dimension $d$, ($1 <d <1.3$) 
these bounds are very significant and comparable to those obtained from 
production rates at high energy colliders. These bounds are also 
similar to those 
obtained by the analysis of the absence of the decay of the low-energy 
solar and reactor neutrinos \cite{Anchordoqui:2007dp}. Both our bounds 
and those obtained in Ref. \cite{Anchordoqui:2007dp} imply that there 
still is a considerable window of the parameter space for possible 
discovery of the unparticle sector at LHC. Note that even though the 
astrophysical constraints are more restrictive in some cases than our limits, 
it is worth pointing out that ours are {\it direct} experimental limits and 
as such they are subject to fewer uncertainties than the astrophysical limits. 
    
One should finally remark that unparticle couplings could change 
neutrino flavor, and may also convert active neutrino flavors into 
sterile ones. Even though such processes may have interesting astrophysical 
implications \cite{Fetter:2002xx}, they are unlikely to provide better bounds 
on unparticle couplings.

\section*{ACKNOWLEDGMENTS}
This work was supported in part by the
U.S. National Science Foundation Grant No.\ PHY-0555231 and in part by
the University of Wisconsin Research Committee with funds granted by
the Wisconsin Alumni Research Foundation. K.O. Ozansoy acknowledges 
support through 
the Scientific and Technical Research Council 
(TUBITAK) BIDEP-2219 grant.

\end{document}